\documentclass{PoS}

\title{GLADIS: GLobal Accretion Disk Instability Simulation}

\ShortTitle{GLADIS Code}

\author{\speaker{Agnieszka Janiuk}\thanks{To the memory of my Mother}\\
  Center for Theoretical Physics\\
  Polish Academy of Sciences\\
  Al. Lotnikow 32/46, 02-668 Warsaw, Poland\\
        E-mail: \email{agnes@cft.edu.pl}}

\abstract{I present the publicly available code GLADIS (GLobal Accretion Disk Instability Simulation) developed in my reserach group over the years 2002-2017.
  It can be freely downloaded and modified by the users via the link from the Astrophysics Source Code Library. The software computes time-dependent evolution of a black hole accretion disk, in one-dimensional, axisymmetric, vertically integrated scheme.  The main applications are to explain the variability of accretion disks that can be subject to radiation-pressure instability. The phenomenon is relevant for fast variable microquasars, as well as for a class of changing-look AGN.
 }

\FullConference{Multifrequency Behaviour of High Energy Cosmic Sources - XIII - MULTIF2019\\
		3-8 June 2019\\
		Palermo, Italy}

\begin{document}

\section{Introduction}

The accreting black holes observed in our Galaxy frequently exhibit the characteristic pattern of X-ray variablity. Their luminosity oscillations
appear on short timescales (tens-hudreds seconds), and the regular outbursts
typically have a profile with a slow rise, and fast decay of the X-ray flux in time.
Two well-known microquasars, GRS 1915+105, and IGR J17091, in some states
are presenting
the variabity pattern called the 'heartbeat state' \cite{Altamirano2011}, because
their oscillation shape resembles the electrocardiogram. It was also shown, that the timing properties of these heartbeats scale with the estimated mass of the black hole.
In other sources, a substructure in between the main outbursts can be found, as they present a substructure and multiple peaks of varying amplitudes
appear in their lightcurves.
Nevertheless, in several sources, such as the black hole candidates GX 339-4,
or XTE 1550-564, the non-linear variability was suggested based on the selected lighturves \cite{JaniukCzerny2011}.

The lightcurves shape describes the time changes in radiation generated in
the black hole accretion flow. This radiation
is supposed to be driven by the energy
dissipated through the viscous stresses within the relatively cold,
soft X-ray emitting thin disk, and possibly transmitted to the hotter,
hard X-ray emitting coronal flow.
But even regardless of the details of the energy dissipation and radiative
process, the traces of non-linear dynamics of the underlying physical process
may be found by studying the observed lightcurves.
As shown for instance by \cite{misra2004}, the method of correlation dimension used in case of GRS 1915+105 for four of its temporal classes indicated that the underlying dynamical mechanism is a low-dimensional chaotic system.
Also, the method of reccurrence analysis used by \cite{sukova2016} revealed
significant traces of nonlinear dynamics in three other sources: GX 339-4,
XTE J1550-564 and GRO J1655-40, particularly in the disk-dominated soft state,
as well as in the intermediate states at the rising and declining phase of the
outburst.

\section{Non-linear processes in accretion disk}

Non-linear variability induced by the accretion disk instability, leads to periodic changes of
brightness, and limit-cycle type of behaviour.

\subsection{Local solutions}

The vertically averaged heating rate locally in the accretion disk, reads:
\begin{equation}
Q_{+} = \frac{3}{2} \alpha P H \Omega_{\rm K}\nonumber
,\end{equation}
where $\alpha$ is the viscosity parameter as introduced by \cite{ss73}, $P$
is the pressure, $H$ is the disk half-thickness, and $\Omega_{K}$ is the Keplerian angular velocity.
Cooling due to advection and radiative losses is given by:
\begin{equation}
Q_{-} = \frac{4 \sigma_{B} T^{4}}{3 \kappa \Sigma}+ F_{\rm tot}(1-f_{adv})\nonumber
\end{equation}
where $\sigma_{B}$ is the Stefan-Boltzmann constant, $\kappa=0.34$ is the electron-scattering opacity, $T$ is the disk mid-plane temperature, and $\Sigma$ is
the disk surface density.

The closing equation for a stationary structure of the disk, is that for the
total energy flux emitted from the surface unit area
\begin{equation}
F_{\rm tot} = {3 G M \dot M \over 8 \pi r^{3}} f(r)\nonumber
\label{eq:ftot}
\end{equation}
Here the global parameters are the mean mass accretion rate, $\dot M$, and black hole mass, $M$. The sero torrque boundary condition for $f(r)$ is typically used.
The advection parameter in general
is related the radial derivatives of density and temperature, but in a stationary model can ba adopted as a constant, $f_{\rm adv}$, on the order of 0.3.
The accretion rate can be convieniently expressed in Eddington units:
\begin{equation}
\dot{m}=\frac{\eta \dot{M}c^{2}}{ L_{\rm Edd}}=\frac{\dot{M}}{\dot M_{\rm Edd}}; ~~~
L_{\rm Edd} = \frac{4 \pi G M_{\rm BH}m_{\rm p}c}{\sigma_{\rm T}}= 1.3\cdot 10^{38} (M_{\rm BH}/M_{\odot})\nonumber
\end{equation}

The only non-vanishing stress-tensor component scales with pressure
as:
\begin{equation}
T_{r \phi} = - \alpha P\nonumber
\label{pirphi}
\end{equation}
where the total pressure $P$ is given by the gas and radiation components:
\begin{equation}
P = P_{\rm gas} + P_{\rm rad} = \frac{\rho k_B}{m_p} T+ \frac{4 \sigma_B}{3 c} T^{4}\nonumber
\end{equation}

\subsection{Instabilities in accretion disks}
In accretion disks, we can have two types of thermal-viscous
instabilities: (i) radiation pressure instability, $T\sim \Sigma^{-1/4}$ \cite{lightman74}, and (ii) partial Hydrogen ionisation instability,
driven by inverse dependence of opacity on temperature
\cite{Smak84}.
The first of these instabilities may lead to
the short term limit cycle oscillations in black hole x-ray
binaries (tens-hundreds seconds scales), or
to a cyclic activity of quasars (scales of tens-thousands of
years).
The second type of instability has proven to induce the X-ray novae eruptions (scales of months-years), while it may also be related to the long-term activity cycles in AGN (scales of millions of
years).

The localisation of the unstable regions in the disk shows two separate regions of the radiation pressure dominated zone, and the zone dominated by the
partial Hydrogen ionisation \cite{JaniukCzerny2011}.
In case of the radiation pressure instability,
the disk can be stabilized by
a strong jet/wind, which takes away some fraction of the locally dissipated energy flux.
Also, by definition, the heating prescription will affect the disk behaviour, and the gass pressure dominated disk is completely stable.
The influence of the companion star can also manifest itself as a stabilizing mechanism, when the mean mass inflow rate drops below the critical value for the instability onset.
Finally, as shown in \cite{janiukmisra2012}, the viscous fluctuations may partly stabilize the accretion disk.

\section{Code GLADIS}
The Code GLADIS (for GLobal Accredion Disk Instability Simulation)
computes the dynamics of the
radiation pressure unstable disk.
There are two governing partial-differential equations to be solved numerically:
\begin{equation}
\frac{\partial \Sigma}{\partial t} = \frac{1}{r} \frac{\partial }{\partial r}( 3 r^{1/2} \frac{\partial }{\partial r}( r^{1/2} \nu \Sigma )) \nonumber
\end{equation}
and
\begin{equation}
\frac{\partial \ln T}{\partial t} + v_r \frac{\partial \ln T}{\partial r} \nonumber
\\
= \frac{4 - 3 \beta }{12 - 10.5 \beta} (\frac{\partial \ln \Sigma}{\partial t} - \frac{\partial \ln H}{\partial t} + v_r \frac{\partial \ln \Sigma}{\partial r}  ) \nonumber
\\
+ \frac{Q_{+} - Q_{-}}{(12-10.5 \beta)PH} \nonumber
\end{equation}
The derivation of above formula can be found e.g. in \cite{PBK81}.

 GLADIS computes the time-dependent evolution of a black hole
accretion disk, in 1 dimensional, axisymmetric, vertically integrated scheme.
The code solves the above equations for surface
density and temperature evolution, i.e. given by
viscous diffussion and energy conservation.

\begin{figure}
\includegraphics[width=0.8\textwidth]{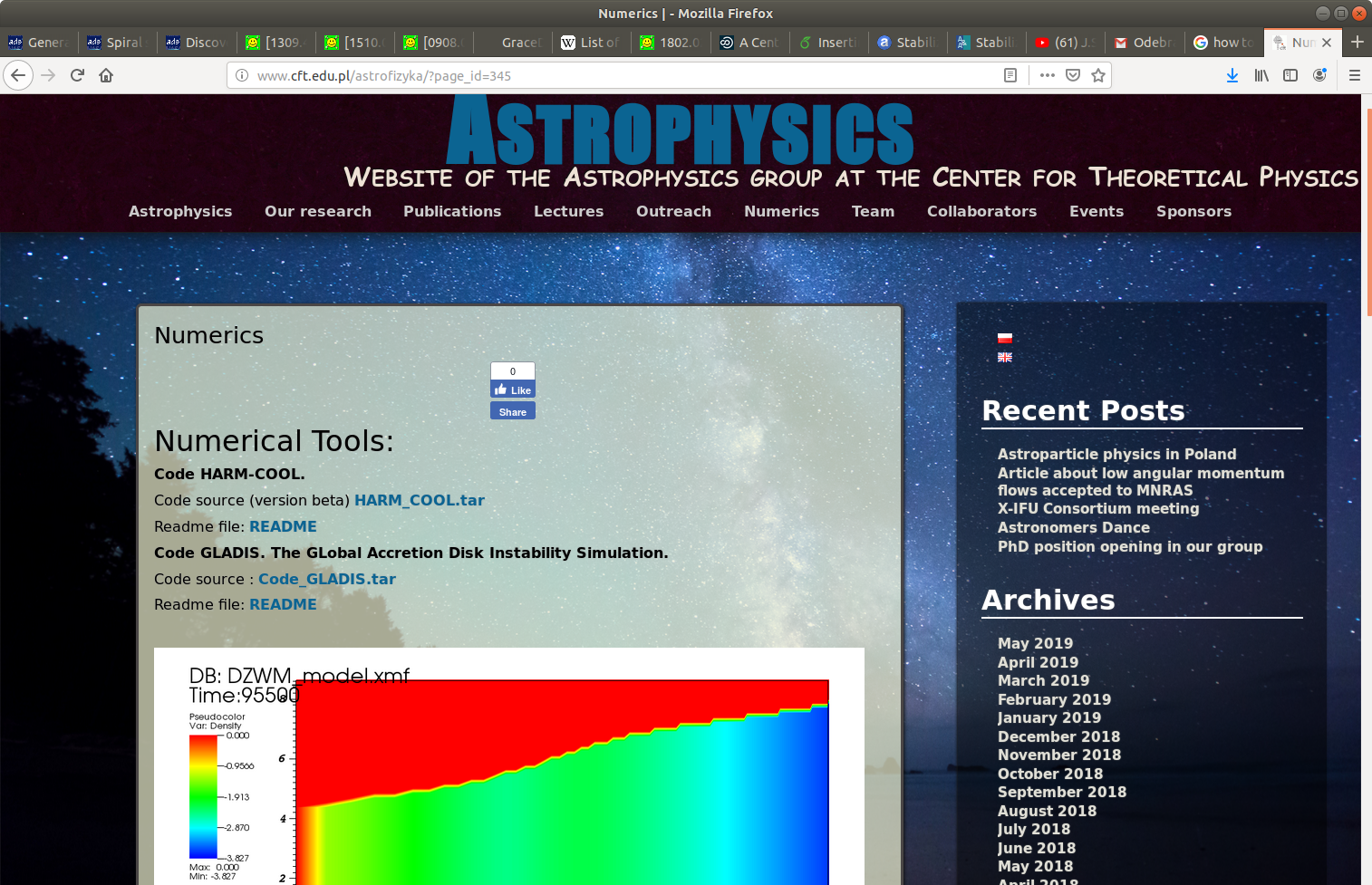}
\caption{Website to download the code GLADIS}
\label{fig:website}
    \end{figure}

The code has been indexed in the Astrophysics Source Code Library \textit{http://ascl.net/1812.002} and ADS database (2018ascl.soft12002J)
The package (tar.gz) with the source code is available at \textit{http://www.cft.edu.pl/astrofizyka/ under tab Numerics}.
The User can download the code sources written in C/C++, a sample makefile, the macros for initial configuration, and a running script.
All the sourecs are available for download from the website of the Astrophysics Group at CTP PAS (see Figure \ref{fig:website}).

The initial conditions assume the stationary disk with a solution numerically integrated with Newton-Raphson scheme for the density, mid-plane temperature, and thickness profiles. The latter is taken from the hydrosytatic equilibrium condition.
The time dependent simulation is run by the Eulerian code, with the $4^{th}$
order Runge-Kutta  and Predictor-Corrector methods.
The code is parallelized with MPI technique and can be run on multiple processors.
The model parameters are stored in the configuration file \emph{dysk-zwm.ini}.
Their values refer to the physical setup as in \cite{janiuk2005}, \cite{janiuk2007}, and \cite{janiukmisra2012}.
The code stores the output in various formats: ASCII and HDF5. Their content is described in detail in the README manual.

\subsection{Exemplary results for the set of parameters reproducing the microquasar IGR J17091}

We run the sample simulation with the set of parameters that represent microquasar IGR J17091-3624.
In this source, the black hole mass is of $M=6 M_{\odot}$, and accretion rate estimated on $\dot M= 0.86-2.21\times 10^{-8} M_{\odot}/yr$.
We use the upper limit value which corresponds
to $\dot m=0.1$ in Eddington units.
The adopted viscosity parameter is $\alpha=0.1$, and the outer radius of the grid was located at $R_{\rm out}=1000 R_{\rm Schw}$.
We also prescribe a moderately stron wind outflow, with the normalisation $A=15$.

\begin{figure}
     \includegraphics[width=0.5\textwidth]{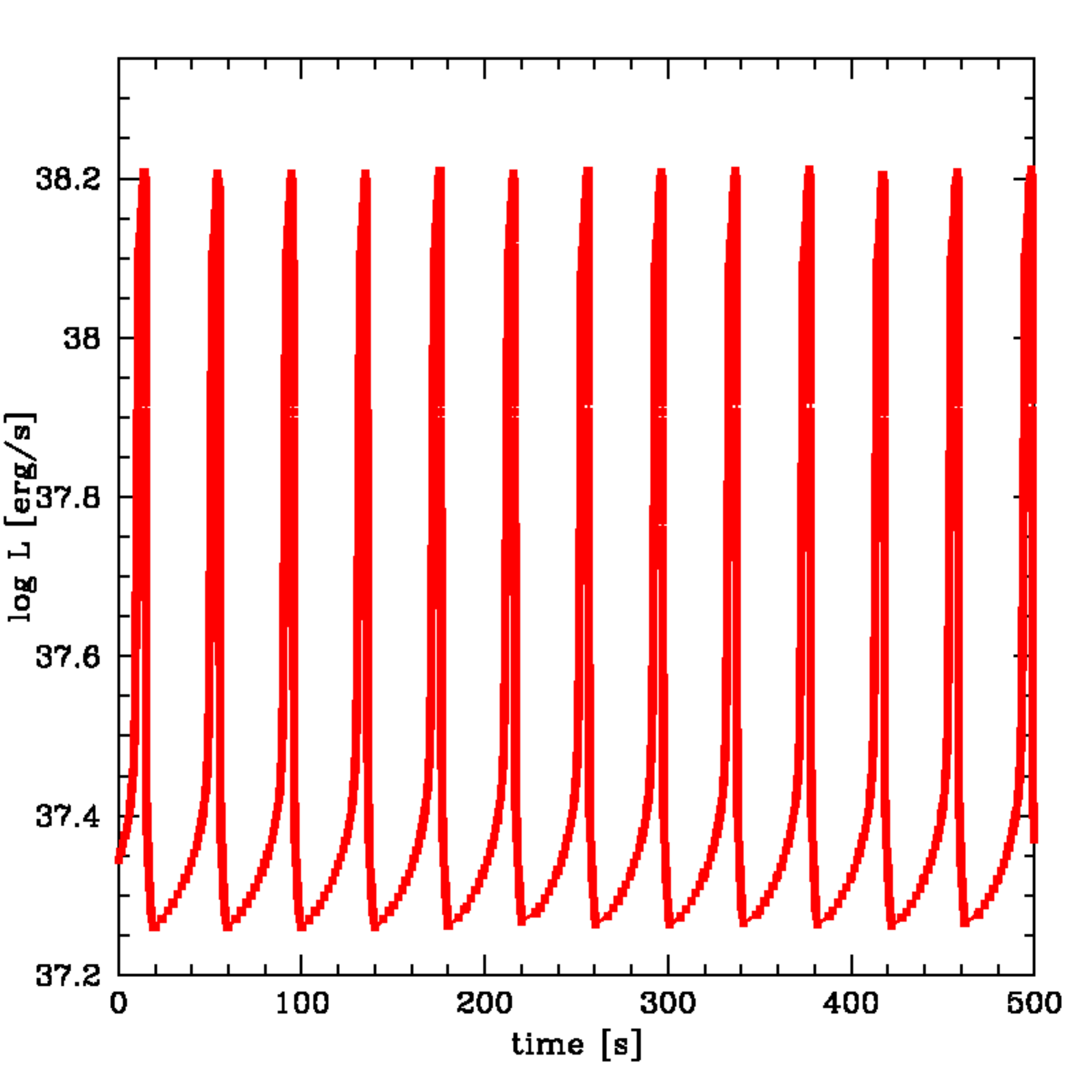}
     \includegraphics[width=0.5\textwidth]{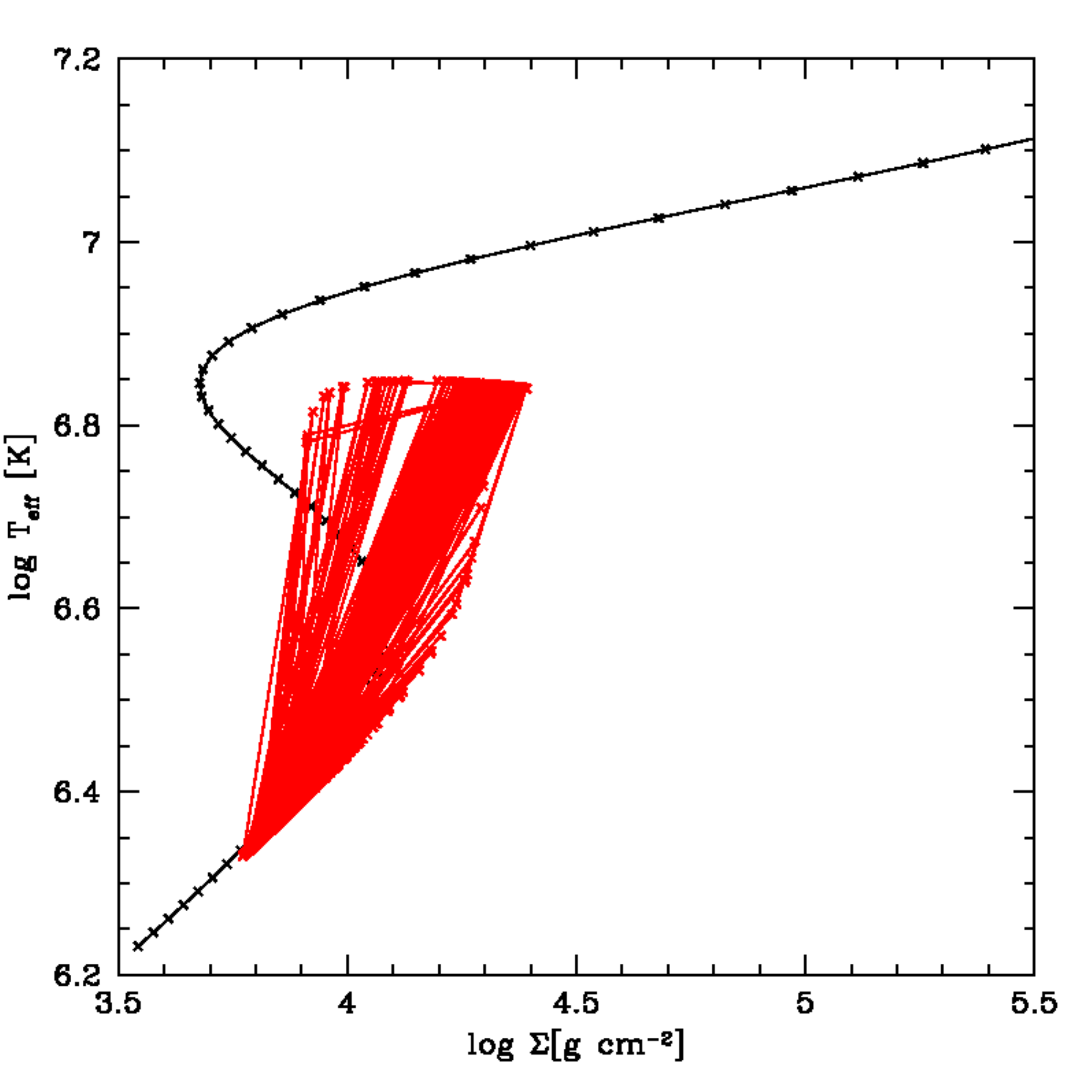}
\caption{Left: Sample lightcurve simulated with the code GLADIS. Right: evolution of surface density and disk temperature at the inner radius. Parameters of the simulation refer to the source IGR 17091.}
\label{fig:IGR_lc}

    \end{figure}

In Figure \ref{fig:IGR_lc} we show a model lightcurve produced with the simulation using GLADIS.
We also show the evolution of the surface density and disk temperature at the inner radius, $r=3.01 R_{\rm Schw}$. It follows the loop around the unstable branch of the stability curve.

The integrated variablity pattern reproduces the heartbeat oscillation as observed in disk-dominated states of the microquasar IGR J17091 (see details in \cite{janiuketal2015}).
As we have shown, the wind launched from the accretion disk and the heartbeat
oscillations manifest actively their role in this source. 
The amplitudes of quasi-periodic flares are anti-correlated with
 the strenght of wind outflow.
Moreover, the wind may help to partially or even completely stabilize the heartbeat
during the soft state, i.e. the oscillations cease, when a stron wind is detected.
We found that the constraints on the wind outflow rate from Chandra data in the
no-heartbeat state  give the mass loss rate in the wind  $\dot M_{\rm wind} = 3-4.7 \times 10^{17}$ g s$^{-1}$, while in the heartbeat state the wind mass loss is ten times smaller.
The CLOUDY photoionisation code results
are consistent with the wind strength and spatial dimension
of the wind launching region determined from the GLADIS model.

\section{Discussion of possible applications of the radiation pressure instability and its modeling with the code GLADIS}

\subsection{Viscosity fluctuations}

Another feature implemented in the GLADIS are the propagating viscosity fluctuations. The idea has been introduced by \cite{lyubarskii} to interpret the nature of accreting sources that exhibit a stochastic variability. Such kind of behaviour may naturally arise through viscous fluctuations in a turbulent disk. 
With this option, the viscosity in accretion disk is no longer a constant, but is a function of radius and time, as given by:
\begin{equation}
\alpha(r,t)=\alpha_{0} [1+\beta(r,t)] \nonumber
\end{equation}
\begin{equation}
\beta_{n}=b_{0}u_{n}=b_{0}(-0.5 u_{n-1} + RAN_{n}) \nonumber
\end{equation}
We have shown that the above viscosity flickering stabilizes the disk
oscillations induced by the radiation pressure \cite{janiukmisra2012}. The amplitude and period of flares decrease with the $b_{0}$ parameter.

\subsection{Modified stress tensor}
The simulations made with
the code GLADIS have been used also as a new method for the black hole mass
determination. In particular, the method was applied to the source HLX-1 \cite{Wu2016}. 
This hyperluminous X-ray source ($10^{42}$ erg s$^{-1}$) near the spiral
galaxy ESO 243-49 underwent recurrent outbursts within a period of  about 400
days, is possibly the best candidate for an intermediate mass black hole (IMBH).The timescales and amplitudes of these outbursts are consistent with the
evolution of the radiation-pressure unstable accretion flow, if we account for
the strong global magnetic field which can stabilize the disk. 
The parameter  $\mu$ introduced to scale the strength of the radiation pressure
contribution to the stress tensor as:
\begin{equation}
   T_{\rm r \phi} = \alpha {P^\mu P_{\rm gas}^{1 - \mu}} \nonumber
\end{equation}
can give an effective 
prescription for the magnetic field \cite{grzedzielski2017a}.

We found, that with $\mu=0.54$ the source HLX-1
fits in the mass-scale relation, spanning several orders of magnitude in
the black hole mass range, from
the two known microquasars, up to the Giga-Hertz Peak Spectrum
radio sources \cite{czerny2009}.
Other best-fit parameters for this source are its black hole mass $M_{BH}=1.5 \times 10^{5} M_{\odot}$, and accretion rate of $\dot m = 0.09-0.18$ of Eddington rate.

Finally, with implementation of the code GLADIS,
we studied the influence on the disk stabilisation made by
   changes of opacities due to the absorption on 
heavy atoms.
It is supposed that they have a local influence on the accretion disk stability,
but this effect is very sensitive to the black hole mass.
Our global simulations, with a range of $\mu$-paramater values for the stress tensor, have shown that  the 
limit cycle oscillations appear. They are suffering some disturbances from the \textit{Iron Opacity Bump}, but nevertheless are still expected in AGN over a wide range of black hole mass \cite{grzedzielski2017b}.


 \section{Summary}
 
\begin{itemize}
\item We have shown that global time-dependent evolution of the accretion disk on viscous timescale
  leads to the regular limit-cycle oscillations for $T_{r\phi}\propto \alpha P$, even at moderate $\dot m$.
  \item Microquasar IGR J17091 
was studied with respect to its unique properties,
and it is the second known source (after GRS 1915+105) that proves the
limit-cycle oscillations. Other sources, such as GX 339-4, and XTE J155,
also show significant tracers of non-linear dynamics. Radiation pressure instability
is universal across the BH mass-scale.
  \item Code GLADIS is now available for the community to download and use. By mens of this tool, the user may search for
    the limit-cycle periodicities in the astrophysics sources.
    \item The most general version of the code setup allows for the wind parametrisation, and also $\alpha$-flickering propagations. Recently the code has been used in a number of applications, published in the specialist literature.
\end{itemize}

{\bf Acknowledgements}
We acknowledge partial support from grant DEC-2016/23/B/ST9/03114
awarded by the Polish National Science Center.

\bigskip
\bigskip
\noindent {\bf DISCUSSION}

\bigskip
\noindent {\bf GALINA LIPUNOVA:} Can the code be changed if one wants to account for a different vertical structure? 

\bigskip
\noindent {\bf AGNIESZKA JANIUK:} The GLADIS code is vertically averaged. Nevertheless, the vertical structure is taken into account via the averaging coefficients. They are defined as the ratio between the vertically integrated density, pressure, and radiation flux, with respect to their equatorial values times the disk thickness (see Eqs. 16-18 in Janiuk et al. 2002).
The coefficients were computed from the model of vertical structure that we had, but they might be verified with a different model and then changed in the GLADIS code.

\bigskip


\begin{thebibliography}{99}
  
\bibitem{Altamirano2011} D. Altamirano, T. Belloni, M. Linares, et al., {\it The Faint "Heartbeats" of IGR J17091-3624: An Exceptional Black Hole Candidate}, \emph{ApJL} {\bf 742} 17 (2011)
\bibitem{czerny2009} B. Czerny, A. Siemiginowska, A. Janiuk, et al., {\it Accretion Disk Model of Short-Timescale Intermittent Activity in Young Radio Sources}, \emph{ApJ} {\bf 698} 840 (2009)
\bibitem{grzedzielski2017a} M. Grzedzielski, A., Janiuk, B. Czerny, Q. Wu, {\it Modified viscosity in accretion disks. Application to Galactic black hole binaries, intermediate mass black holes, and active galactic nuclei } \emph{A\&A} {\bf 603} 110 (2017a)
\bibitem{grzedzielski2017b} M. Grzedzielski, A., Janiuk, B. Czerny,
  {\it Local Stability and Global Instability in Iron-opaque Disks} \emph{ApJ} {\bf 845} 20 (2017b)
\bibitem{janiuk2005} A. Janiuk, B. Czerny, {\it Time-delays between the soft and hard X-ray bands in GRS 1915+105}, \emph{MNRAS} {\bf 356} 205 (2005)
  \bibitem{janiuk2007}  A. Janiuk, B. Czerny, {\it Accreting corona model of the X-ray variability in soft state X-ray binaries and active galactic nuclei}, \emph{A\&A} {\bf 466} 793 (2007) 
  \bibitem{JaniukCzerny2011} A. Janiuk, B. Czerny, {\it On different types of instabilities in black hole accretion discs: implications for X-ray binaries and active galactic nuclei}, \emph{MNRAS}, {\bf 414} 2186 (2011)
  \bibitem{janiuketal2015} A. Janiuk, M. Grzedzielski, F. Capitanio, S. Bianchi,
   {\it Interplay between heartbeat oscillations and wind outflow in microquasar IGR J17091-3624}, \emph{A\&A} {\bf 574} 92 (2015)
\bibitem{janiukmisra2012} A. Janiuk, R. Misra, {\it Stabilization of radiation pressure dominated accretion disks through viscous fluctuations}, \emph{A\&A} {\bf 540} 114 (2012)
\bibitem{lightman74} A.P. Lightman, D.M. Eardley, {\it Black Holes in Binary Systems: Instability of Disk Accretion}, \emph{ApJ} {\bf 187} L1 (1974)
\bibitem{lyubarskii} Y.E. Lyubarskii, {\it Flicker noise in accretion discs},
  \emph{MNRAS}, {\bf 292} 679 (1997)
\bibitem{misra2004} R. Misra, K.P. Harikrishnan, B. Mukhopadhyay, G. Ambika, A.K. Kembhavi, {\it The Chaotic Behavior of the Black Hole System GRS 1915+105},
  \emph{ApJ} {\bf 609} 313 (2004)
  \bibitem{PBK81} B. Paczynski, G. Bisnovatyi-Kogan, {\it A Model of a Thin Accretion Disk around a Black Hole}, {\emph Acta Astron.} {\bf 31} 283 (1981)
\bibitem{sukova2016} P. Sukova, A. Janiuk, M. Grzedzielski, {\it Chaotic and stochastic processes in the accretion flows of the black hole X-ray binaries revealed by recurrence analysis}, \emph{A\&A} {\bf 586} 143 (2016)
  \bibitem{ss73} N.I. Shakura, R.A. Sunyaev, {\it Black holes in binary systems. Observational appearance.}, {\emph A\&A} {\bf 24} 337 (1973)
  \bibitem{Smak84} J. Smak, {\it Accretion in cataclysmic binaries. IV. Accretion disks in dwarf novae.}, {\emph Acta Astron.} {\bf 34} 161 (1984)
  \bibitem{Wu2016} Q. Wu, B. Czerny, M. Grzedzielski, et al., {\it The Universal “Heartbeat” Oscillations in Black Hole Systems Across the Mass-scale}, \emph{ApJ} {\bf 833} 79 (2016)
\end{thebibliography}
\end{document}